\def\C(#1){|#1|}

\def\eqar#1{\begin{eqnarray} #1 \end{eqnarray}}
\def\eq#1{\begin{equation} #1 \end{equation}}
\def\eqn#1{\begin{equation} \nonumber #1 \end{equation}}

\def\mbit#1{\mbox{\boldmath$ #1 $}}
\def\Def{\buildrel \Delta \over =}

\def\Ind(#1){\Delta\left[#1\right]}

\def\Inner(#1,#2){\langle #1, #2 \rangle}

\documentclass[10pt,conference]{IEEEtran}
\usepackage{graphicx}
\usepackage{amsmath}
\usepackage{amsfonts}
\usepackage{array}
\newtheorem{thm}{Theorem}
\newtheorem{cor}[thm]{Corollary}
\newtheorem{lem}[thm]{Lemma}

\newtheorem{ex}[thm]{Example}
\newtheorem{defn}[thm]{Definition}

\begin{document}

\title{The Partition Weight Enumerator of MDS Codes and
its Applications.}
\author{Mostafa El-Khamy
$^*$ and  Robert J. McEliece$^{**}$\\
\begin{tabular}{c}
Electrical Engineering Department\\
California Institute of Technology\\
Pasadena, CA 91125, USA\\
\begin{tabular}{c c}
$^*$E-mail: mostafa@systems.caltech.edu & $^{**}$E-mail: rjm@systems.caltech.edu \\
\end{tabular}
\end{tabular}}
\maketitle

\begin{abstract}
A closed form formula of the partition weight enumerator of maximum
distance separable (MDS) codes is derived for an arbitrary number of
partitions. Using this result, some properties of MDS codes are
discussed. The results are extended for the average binary image of
MDS codes in finite fields of characteristic two. As an application,
we study the multiuser error probability of Reed Solomon codes.
\end{abstract}

\section{Introduction and Summary}

In this paper, we introduce a generalized weight enumerator, which
we call the partition weight enumerator (PWE).  Our main result is a
simple closed-form expression for the PWE of an arbitrary MDS, e.g.,
Reed-Solomon (RS), code (Theorem \ref{McPart}). This generalizes the
results of Kasami et al. \cite{kasami97} on the split weight
enumerator of RS codes.

We then derive weight enumerators for the average binary image of MDS (Reed-Solomon) codes defined
over finite fields of characteristic two (Section \ref{secABPWE}).

We also derive a strong symmetry property for MDS codes (Theorem
\ref{wghtMDS}) which allows us to obtain improved bounds on the
decoder error probability for RS codes (Section VI).

Finally, we discuss possible applications of the PWE, including the
analysis of the performance of RS codes in a multiuser setting
(Section VII).

\section{Preliminaries \label{pre}}

We begin by generalizing the notion of Hamming weight. Let $V_n(F_q)$
denote the vectors of length $n$ over the finite field of $q$ elements $F_q$.  Suppose the
coordinate set $N= \{1,2,\ldots,n\}$ is partitioned into $p$
disjoint subsets  $N_1, \ldots, N_p$, with $|N_i| = n_i$, for $i =
1, \ldots, p$.  Denoting  this partition  by $\mathcal{T}$, the
$\mathcal{T}$-weight profile of a vector $\mbit{v} \in V_n(F_q)$ is
defined as $\mathcal{W}_{\mathcal{T}}(\mbit{v})= (w_1, \ldots,w_p)$, where
$w_i$ is the Hamming weight of $\mbit{v}$ restricted to $N_i$.
Now we generalize the notion of code weight enumerator. Given a
code $\mathbb{C}$ of length $n$, and an $(n_1,
 n_2,...,n_p)$ partition $\mathcal{T}$ of the $n$ coordinates of $\mathbb{C}$,
  the $\mathcal{T}$-weight enumerator of $\mathbb{C}$ is the set of numbers
\eqn{A^{\mathcal{T}}(w_1, \ldots, w_p) =
 |\{\mbit{c} \in \mathbb{C}: \mathcal{W}_{\mathcal{T}}(\mbit{c}) = (w_1, \ldots,w_p)\}|.}

The weight enumerator of $\mathbb{C}$ is \eq{E_{\mathbb{C}}(w)=|\{\mbit{c} \in \mathbb{C}:
\mathcal{W}(\mbit{c})=w\}|,} where $\mathcal{W}(\mbit{c})$ is the Hamming weight of $\mbit{c}$.
The weight generating function (WGF) of
$\mathbb{C}$ is the polynomial
 $\mathbb{E}_{\mathbb{C}}(\mathcal{X})=\sum_{h=0}^{n} E_{\mathbb{C}}(h) \mathcal{X}^h$.
 (The subscript $\mathbb{C}$ may be dropped when there is no ambiguity
about the code.) For an $(n,k,d)$ MDS code over $F_q$, the minimum distance is $d=n-k+1$
\cite{Mc02book} and
  the weight distribution is given by  \cite[Th.
  25.7]{VLintWilson01} for weights $i \geq d$,
\eqar{ \label{Ai1} E(i) &=& {n \choose i} \sum_{j=d}^i {i \choose j}
(-1)^{i-j} (q^{j-d+1}-1).} The \emph{partition weight generating
function} (PWGF) is
\eqar{\label{par}\mathbb{P}^{\mathcal{T}}_{\mathbb{C}}(\mathcal{X}_1,...,\mathcal{X}_p)=
  \sum_{w_1=0}^{n_1}...\sum_{w_p=0}^{n_p}
A^{\mathcal{T}}(w_1,...,w_p)
\mathcal{X}_1^{w_1}...\mathcal{X}_p^{w_p}.}
For the special case of $p=2$, $A^{\mathcal{T}}(w_1,w_2)$ is termed
the\emph{ split weight enumerator} in the literature \cite{McSln}.
The \emph{input-redundancy weight enumerator} (IRWE), $R(w_1,w_2)$,
is the number of codewords with input weight (weight of the
information vector) $w_1$ and redundancy weight $w_2$. For a
systematic code, if $\mathcal{T}$ is an $(k, n-k)$ partition such
that the first partition constitutes of the coordinates of the
information symbols, then $R(w_1,w_2)=A^{\mathcal{T}}(w_1,w_2).$ The
\emph{input-output weight enumerator} (IOWE) $O(w,h)$ enumerates the
codewords of total Hamming weight $h$ and input weight $w$. Assuming
that the first partition constitutes of the information symbols,
then $O(w,h)=R(w,h-w)$. For an $(k, n-k)$ partition $\mathcal{T}$,
it is straight forward that \eq{E(h)=\sum_{w=0}^{k}
A^{\mathcal{T}}(w,h-w)=\sum_{w=0}^k O(w,h).}
 The IOWE and IRWE are used in the literature to study
 the bit error probabilities of codes (e.g. \cite{Ben98}).

 For a systematic code, let the
$j$th partition constitute of information symbols, then the $j$th
IOWE, \eq{\label{Oj} O^{j}(w,h) =|\{\mbit{c} \in \mathbb{C}:
  (\mathcal{W}(N_j)=w) \wedge (\mathcal{W}(\mbit{c})=h) \}|,}
 is the coefficient of
$\mathcal{X}^{w}\mathcal{Y}^{h}$ in
$\mathbb{O}^{j}(\mathcal{X},\mathcal{Y})=
\mathbb{P}^{\mathcal{T}}_{\mathbb{C}}(\mathcal{Y},\mathcal{Y},.,
\mathcal{XY},.,\mathcal{Y})$ where the $\mathcal{X}_i$s in
(\ref{par}) are substituted by $\mathcal{X}_i \Rightarrow
\mathcal{Y}$ if $i\neq j$ and $\mathcal{X}_i \Rightarrow \mathcal{X
Y}$ if $i=j$.
\section{Partition weight enumerator of MDS codes}

\thm{ \label{Th:Part1} For a $p$-partition $\mathcal{T}$, the PWE of
an $(n,k,d)$ MDS code $\mathbb{C}$ over $F_q$,
$A^{\mathcal{T}}(w_1,w_2,...,w_p)$, is given by \eqar{\label{Part}
\nonumber {n_1 \choose w_1}....{n_p \choose w_p} \sum_{j_1=0}^{w_1}
{w_1 \choose j_1} (-1)^{w_1-j_1}
\sum_{j_2=0}^{w_2} {w_2 \choose j_2} (-1)^{w_2-j_2}\\
\nonumber .... \sum_{j_p=d-\sum_{z=1}^{p-1}j_z} ^{w_p}{w_p \choose
j_p} (-1)^{w_p-j_p} (q^{k-n+\sum_{z=1}^{p}j_z}-1).}}

\emph{Sketch of Proof:} Let $R_i$ be a subset of $N_i$ for
$i=1,2,..p$. Define $S(\mbit{c})$ to be the support set of the
codeword $\mbit{c}$ and $f(R_1,..,R_p)\Def |\mbit{c}: \{S(\mbit{c})
\cap N_i\}= R_i, \;\; \forall \; i|.$ Let $S_i \subseteq N_i$, then
from the MDS property of $\mathbb{C}$, we have \eqar{\nonumber
g(S_1,...,S_p) \Def \sum_{R_1 \subseteq S_1} ...\sum_{R_p \subseteq
S_p}f(R_1,...,R_p)\\\label{Th1:core}
 =\left\{
\begin{array}{ll}
    1, & \sum_{i=1}^p|S_i|< d; \\
    q^{1-d+\sum_{i=1}^p|S_i|}, & n \geq \sum_{i=1}^p|S_i| \geq d. \\
\end{array}%
\right.} Successively applying M\"{o}bius Inversion \cite[Th.
25.1]{VLintWilson01}, and observing that the PWE is equal to
\eqn{A^{\mathcal{T}}(w_1,...,w_p) = \prod_{i=1}^p\left(\sum_{R_i
\subseteq N_i,|R_i|=w_i}\right)  f(R_1,...,R_p),} the result
follows.
  \hfill \QED
\lem{ \label{lem:split}Let $\mathcal{T}$ be an $(n_1,n_2)$
partition, then $A^{\mathcal{T}}(w_1,w_2) = E(w_1+w_2) \frac{ {n_1
\choose w_1} {n_2 \choose w_2}}{{n \choose {w_1+w_2}}}.$}

   \emph{Sketch of
Proof:} From Th. \ref{Th:Part1}, the split weight enumerator is
\eqar{\nonumber \label{IRWERS} A^{\mathcal{T}}(w_1,w_2) =  {n_1
\choose w_1}{n_2 \choose w_2} \sum_{j=0}^{w_1} {w_1 \choose j}
(-1)^{w_1-j}\\\sum_{i=d-j}^{w_2}{w_2 \choose i} (-1)^{w_2-i}
(q^{i+j-d+1}-1).} By changing the order of the summations, doing a
change of variables and comparing with (\ref{Ai1}), we are done.
\hfill \QED

The PWE of MDS codes does not depend on the orientation of the
coordinates with respect to the partitions but only on the
partitions' sizes and weights (see (\ref{Th1:core})). Thus the ratio
of $A^{\mathcal{T}}(w_1,w_2,...,w_p)$ to $E\left(\sum_{i=1}^p
w_i\right)$ is the probability that the nonzero symbols are
distributed among the partitions with a $\mathcal{T}$-profile
$(w_1,w_2,...,w_p)$, i.e., \thm{\label{McPart} For an $(n,k,d)$ MDS
code $\mathbb{C}$ the $p$-partition weight enumerator is given by
\eqar{\label{PartMc} \nonumber A^{\mathcal{T}}(w_1,w_2,...,w_p)=
E(w) \frac{ {n_1 \choose w_1} {n_2 \choose w_2} ....{n_p \choose
w_p}}{{n \choose w}},} where $w=\sum_{i=1}^p w_i$ and
$E(w)=|\{\mbit{c} \in \mathbb{C}: \mathcal{W}(\mbit{c})=w\}|$.}

The proof of Th. \ref{McPart} also follows by generalizing the
proof of Lem. \ref{lem:split} to any number of partitions.
\cor{The IOWE of a systematic MDS code, is $O(w,h)=E(h) \frac{{k
\choose w}{{n-k} \choose {h-w}}}{{n \choose h}}$ for $h \geq d$.}

Since $\sum_{w}O(w,h)=E(h)$ and ${n \choose h}=\sum_{w=0}^k{k
\choose w}{{n-k} \choose {h-w}}$, we have proved this interesting
identity (using (\ref{Ai1}) and (\ref{IRWERS}))
\eq{\label{id1} \sum_{w=0}^k {k \choose w}{{n-k} \choose
{h-w}} \Psi(w)= \Psi(0)\sum_{w=0}^k{k \choose w}{{n-k} \choose
{h-w}},}
where $g(h,w,i)\Def {h-w \choose i} (-1)^{h-w-i}$ and
 \eqn{\Psi(w)\Def \sum_{j=0}^{w} {w
\choose j} (-1)^{w-j} \sum_{i=d-j}^{h-w}g(h,w,i) (q^{i+j-d+1}-1).}
\cor{For an MDS code of length $n$, the number of codewords which
are zero at a fixed subset of coordinates of cardinality $n-h$ and
are nonzero in the remaining $h$ positions is $\frac{E(h)}{{n
\choose h}}$.}

\emph{Proof:} Let $\mathcal{T}$ be the implied $(h,n-h)$ partition, then the
required number of codewords is $A^{\mathcal{T}}(h,0)$ (See Lem.
\ref{lem:split}.) \hfill \QED
 \ex{ {The PWGF for the $(1,1,2,3)$ partition of
the coordinates of the $(7,3,5)$ RS code over $F_8$ is

$\mathbb{P}(\mathcal{V,X,Y,Z})=1+21 \mathcal{V X Y}^2 \mathcal{Z}$
$+ 42 \mathcal{V X Y Z}^2 +
   21 \mathcal{V Y}^2 \mathcal{Z}^2 + 21 \mathcal{X Y}^2 \mathcal{Z}^2 +
   63 \mathcal{V X Y}^2 \mathcal{Z}^2$
     $+ 7 \mathcal{V X Z}^3 +
    14 \mathcal{V Y Z}^3$ $+14 \mathcal{X Y Z}^3 + 42 \mathcal{V X Y Z}^3$ $+
    7 \mathcal{Y}^2 \mathcal{Z}^3 +
    21 \mathcal{V Y}^2 \mathcal{Z}^3  +21\mathcal{ X Y}^2 Z^3 +
     217 \mathcal{V X Y}^2 \mathcal{Z}^3.$}

    It could be checked that the sum of the coefficients is $8^3$.

\section{\label{secABPWE} Average Binary Partition Weight Enumerator of MDS Codes}
 The binary image $\mathbb{C}^b$ of an $(n,k)$ code $\mathbb{C}$ over $F_{2^m}$
 is obtained by representing each symbol by an $m$-dimensional
 binary vector in terms of a basis of the field \cite{Mc02book}.
  The weight enumerator of $\mathbb{C}^b$ will vary according to the
 basis used. For performance analysis, one could average the
performance over all possible binary representations of
$\mathbb{C}$. Assuming that the distribution of the bits in the
non-zero symbol follows a binomial distribution, the \emph{average
binary} WGF,
$\tilde{\mathbb{E}}_{{\mathbb{C}^b}}(\mathcal{X})=\sum_{h=0}^{n m}
\tilde{E}(h) \mathcal{X}^{h}$, could be shown to be
\cite{ELKMcAller, Retter91},
\eq{\tilde{\mathbb{E}}_{{\mathbb{C}^b}}(\mathcal{X})= \sum_{h=0}^{n}
\frac{E(h)}{(2^m-1)^h} \left( (1+\mathcal{X})^m-1 \right)^h.}
 In \cite{ELKMcAller}, it was shown that the average binary weight enumerator
approaches that of a normalized binomial distribution for all weights greater
than the average binary minimum distance, $\tilde{d}^b$, of the code
\eq{\tilde{E}(h) \approx q^{-(n-k)}{{m n}\choose{h}}\;\; ;h\geq \tilde{d}^b.}
 Consequently, lower bounds on the
average binary minimum distance were derived \cite{ELKMcAller}.

The \emph{average binary} PWGF gives the average number of
codewords with a specific profile of Hamming weights in the binary
images of the specified partitions.
 \thm{\label{binParthm}
Let $\mathbb{P}^{\mathcal{T}}_{\mathbb{C}}(\mathcal{X}_1,...,\mathcal{X}_p)$ be
the PWGF of an $(n,k)$ code $\mathbb{C}$ over $F_{2^m}$, and
$\mathcal{T}_b$ be the partition of the coordinates of
$\mathbb{C}^b$ induced by $\mathcal{T}$ when the symbols are
represented by bits. Given that $F(\mathcal{Z})=\frac{1}{2^m-1}
((1+\mathcal{Z})^m-1)$, the averaged PWGF of $\mathbb{C}^b$ is
$\tilde{\mathbb{P}}^{\mathcal{T}_b}_{\mathbb{C}^b}
 (\mathcal{Z}_1,...,\mathcal{Z}_p) =
 \mathbb{P}^{\mathcal{T}}_{\mathbb{C}}(F(\mathcal{Z}_1),...,F(\mathcal{Z}_p)).$

\emph{Sketch of Proof:} Assuming a binomial distribution of the bits
in a nonzero symbol, the binary WGF of a partition of symbol weight $w_j$
is $\left(\frac{1}{2^m-1} \sum_{i=1}^m {m \choose i} \mathcal{Z}_j^i
\right)^{w_j}$. If the $\mathcal{T}$-profile of a codeword is
$(w_1,w_2,...,w_j)$, then its WGF is $\prod_{j=1}^{p}
(F(\mathcal{Z}))^{w_j}$. By multiplying with the number of such
codewords, $A^{\mathcal{T}}(w_1,w_2,...,w_p)$, the result follows.
\hfill \QED

The\emph{ average binary} IOWE $\tilde{O}(w_b,h_b)$ enumerates the
codewords with an input weight $w_b$ and an output weights $h_b$ in the average binary image.
\cor{ \label{binIO} Let $\mathcal{T}$ be an $(s,n-s)$ partition
of the coordinates of $\mathbb{C}$ and $O_{\mathbb{C}}(w,h)$
be the corresponding IOWE, then
 the averaged IOWE of $\mathbb{C}^b$ for the partition $\mathcal{T}_b$ is given by
 \eqar{\nonumber \tilde{O}_{\mathbb{C}^b}(w_b,h_b)=\sum_{w=0}^s
\sum_{h=w}^n \frac{O_{\mathbb{C}}(w,h)}{(2^m-1)^h}
 \left( \sum_{j=0}^{h-w} (-1)^{h-w-j}\right.\\
\left. { {h-w} \choose j} {j m \choose h_b-w_b} \right)
  \left( \sum_{j=0}^w (-1)^{w-j} {w \choose j}
 {j m \choose w_b} \right).}}

The proof follows by some algebra \cite{ELKMcPart}.

\section{\label{Coor}
 A relationship between Coordinate weight and the codeword weight.}
Define $\mathbb{C}_h \Def \{\mbit{c} \in \mathbb{C}: \mathcal{W}(\mbit{c})=h\}$. %
We prove an important property of MDS codes in the following lemma.
 \lem{ \label{w1} For an $(n,k,d)$ MDS code $\mathbb{C}$, the total
Hamming weight of any coordinate, summed over all codewords in
$\mathbb{C}_h$, is equal to $\frac{h E(h)}{n}$, where $\mathbb{C}_h$ is the set of
codewords
 of $\mathbb{C}$ with Hamming weight $h$.}

\emph{Sketch of Proof:} Let $\mathcal{T}$ be an $(1,n-1)$ partition,
the required number of codewords is $A^{\mathcal{T}}(1,h-1)$. (See Lem. \ref{lem:split}.) \hfill
\QED

 Since the PWE does not depend on the orientation of the
 coordinates, we have the following theorem,
\thm{\label{wghtMDS} For an $(n,k,d)$ MDS code $\mathbb{C}$, the
ratio of the total weight of any $s$ coordinates of $\mathbb{C}_h$
to the total weight of $\mathbb{C}_h$ is $\frac{s}{n}$. If the $s$
coordinates are `input' coordinates, then $\sum_{w=1}^s {w
O(w,h)}=\frac{s}{n} h E(h)$ for all Hamming weights $h$.}

As a side result, we have proven this identity (c.f. (\ref{id1}))
\eqn{\sum_{w=1}^s {{s-1} \choose {w-1}}{{n-s} \choose {h-w}}
\Psi(w)=\Psi(0)\sum_{w=1}^s{{s-1} \choose {w-1}}{{n-s} \choose
{h-w}}.}
\defn{An $(n,k)$ code $\mathbb{C}$ (not necessary MDS) is said to
have property $\mathcal{A}$, if it satisfies Th. \ref{wghtMDS} for all $s$ and $h$.}

Observe that Th. \ref{wghtMDS} is not true for all linear codes.
For example, the $(5,3)$ binary code defined by the generator matrix%
\eqn{G=\left(\begin{array}{ccccc}
  1 &0  & 0&1 &1 \\
  0 &1  & 0&0  &1 \\
  0 &0  &1 & 0 &1 \\
\end{array}%
\right)} is composed of the $8$ codewords, $\{ 00000$, $10011$, $01001$,
$11010$, $00101$, $10110$, $01100$, and $11111\}$, and doesn't have property
$\mathcal{A}$. (Let the input partition be composed of the first $3$
coordinates.)

It is to be noted that all cyclic codes have property
$\mathcal{A}.$ This is partially justified by the fact that any
cyclic shift of a codeword of weight $h$ is also a codeword of
weight $h$ with $h/n$ of the coordinates holding non-zero elements.
However, this neither implies Th. \ref{wghtMDS} nor is it implied by
Th. \ref{wghtMDS}. ( An extended RS code is an MDS code but not a
cyclic code while an $(7,4)$ binary Hamming code is cyclic but not
MDS.) Also, if a code satisfies property $\mathcal{A}$, it is not
necessary that the code is cyclic or MDS. For example, the first
order Reed Muller codes \cite{VLintWilson01} as well as their duals,
the extended Hamming codes \cite{McSln},
 have property $\mathcal{A}$ but are neither cyclic nor MDS;
 \thm{\label{RM} The first order Reed Muller codes have property $\mathcal{A}$.}

\emph{Proof:} By construction from Hadamard matrices \cite{ELKMcPart}.\hfill
\QED

 In fact, we prove here that if a linear code has property $\mathcal{A}$
 then its dual has property $\mathcal{A}$. This result also
 strengthens Th. \ref{wghtMDS}.
We will start by the MacWilliams identity relating the PWE of a code
with that of the dual code.
\thm{ \label{MacW}Let $\mathbb{C}$ be an
$(n,k)$ linear code over $F_q$ and $\mathbb{C}^{\bot}$ be its dual
code. If $\mathcal{T}$ is
 an $(n1,n2)$ partition of their coordinates, $A(\alpha,\beta)$ and $A^{\bot}(\alpha,\beta)$
are the PWEs of $\mathbb{C}$ and $\mathbb{C}^{\bot}$ respectively,
then $A(\alpha,\beta)$ and $A^{\bot}(\alpha,\beta)$ are related by
\eqn{A^{\bot}(\alpha,\beta)= \frac{1}{|\mathbb{C}|}
\sum_{v=0}^{n_2}\sum_{w=0}^{n_1}A(w,v) \mathcal{K}_{\alpha}(w,n_1)
\mathcal{K}_{\beta}(v,n_2),} such that the Krawtchouk polynomial is
 $\mathcal{K}_{\beta}(v,\gamma)=
\sum_{j=0}^{\beta} {{\gamma -v} \choose {\beta -j}} {v \choose j}
(-1)^j (q-1)^{\beta -j}$ for $\beta=0,1,...,\gamma$.}

\emph{Proof:}  By a straight forward manipulation
 of the Macwilliams identity for the split weight enumerator
\cite[Ch. 5, Eq. 52]{McSln}, \cite{Hwang81}. \hfill \QED.

Define $A_j(\alpha,\beta)$ and $A^{\bot}_j(\alpha,\beta)$ to be the
PWEs
 of $\mathbb{C}$ and $\mathbb{C}^{\bot}$ respectively for an $(1,n-1)$ partition
 of their coordinates
 such that
 the first partition is composed of the $j$th symbol.
\thm{An $(n,k)$ linear code over $F_q$ has property $\mathcal{A}$
iff its dual has property $\mathcal{A}$.}

\emph{Sketch of Proof:} From Th. \ref{MacW}, the PWE of
$\mathbb{C}^{\bot}$ is \eq{ A^{\bot}_i(1,\beta) =
\frac{1}{|\mathbb{C}|} \sum_{v=0}^{n-1}\sum_{w=0}^{1}A_i(w,v)
\mathcal{K}_{1}(w,1) \mathcal{K}_{\beta}(v,n-1).} Since $\mathbb{C}$
has property $\mathcal{A}$, then $A_i(1,v)$ and $A_i(0,v)$ don't
depend on the choice of the coordinate $i$. Counting the total
weight of the codewords in  $\mathbb{C}^{\bot}_{\beta+1}$ by two
different ways, we get $\sum_{i=1}^{n} A^{\bot}_i(1,\beta)
=(\beta+1) E_{\mathbb{C}^{\bot}}(\beta+1)$ (c.f Lem. \ref{w1}).
The
converse follows from that if $\mathbb{C}^{\bot}$ has property
$\mathcal{A}$ then $(\mathbb{C}^{\bot})^{\bot}=\mathbb{C}$ has
property $\mathcal{A}$.
 \hfill \QED
\cor{The extended Hamming codes have property $\mathcal{A}$.}

A similar property holds for the
binary image of MDS codes defined over $F_{2^m}$.
\thm{ \label{wghtbit} Let
$\mathbb{C}$ be an MDS code over $F_{2^m}$  with property
$\mathcal{A}$. 
If $\tilde{O}(w,h)$ is the IOWE of $\mathbb{C}^b$,
where the partition of the coordinates of $\mathbb{C}^b$ is induced
by an $(s,n-s)$ partition of the coordinates of $\mathbb{C}$, then
$\sum_{w_b=1}^{m s} w_b \tilde{O}(w_b,h_b) = \frac{s}{n} h_b
\tilde{E}(h_b)$. }

\emph{Sketch of Proof:} Let $s=1$. Since $\mathbb{C}$ has property
$\mathcal{A}$, then $O(1,h)=\frac{h}{n} E(h)$.
 One can show that
 $\sum_{w_b=1}^{m } w_b \tilde{O}(w_b,h_b)=\frac{h_b}{n}
 \tilde{E}(h_b)$ (See Cor.
\ref{binIO}) by some algebraic manipulations.  \hfill \QED

\section{\label{secsymbit} Symbol and Bit Error Probabilities of RS codes}
In this section, we discuss the application of the PWE in
determining the symbol or bit error probability when systematic RS
codes are used for transmission. (Maximum likelihood (ML) decoding of binary linear codes achieves
the
least bit error probability when the code is systematic \cite{FosLinR98}.)

The codeword error probability (CEP)
is the probability that the received word lies in the
decoding sphere of a codeword other than the transmitted word. The CEP for an $(n,k,d)$ RS code
is determined by
the weight enumerator
of the code and the signal to noise ratio $\gamma$ and is given by \cite{McSw86}
 \cite[Eq. 10-9:20]{Wicker95}
\eq{\label{BMDerr} \Phi_C(\gamma)= \sum_{h=d}^{n} E(h)
\sum_{t=0}^{\tau}P_t^h(\gamma),} where $P_t^h(\gamma)$ is the probability that a
received word is exactly Hamming distance $t$ from a codeword of
weight $h$ and $\tau=\lfloor (d-1)/2\rfloor$ is the Hamming decoding radius.

It is well known that the symbol error probability (SEP)
$\Phi_S(\gamma)$ is derived from $\Phi_C(\gamma)$ by substituting
$E(h)$ with $O_h=\sum_{w=1}^k \frac{w}{k} O(w,h),$  (e.g., \cite[Eq.
10-14]{Wicker95}). From Th. \ref{wghtMDS}, the common approximation
$O_h\approx\frac{h}{n}E(h)$ is exact and
 \eqn{\label{symcod}\Phi_S(\gamma)
=\Phi_C(\gamma)\left|_{E(h)\Rightarrow O_h}\right.=\sum_{h=d}^{n} \frac{h}{n}E(h)
\sum_{t=0}^{\tau}P_t^h(\gamma).}

In case the binary image of an RS code is transmitted, tight bounds
on the CEP of the optimum ML decoder are obtained by using the
average binary weight enumerator in conjunction with well-known
bounds \cite{ELKMcAller}. In case of hard-decision ML decoding of
binary linear codes over an additive white Gaussian noise (AWGN)
channel, the Poltyrev bound for binary symmetric channels
\cite{Polt94} is a tight upper bound. Tight bounds on the CEP of
soft-decision ML decoding of binary linear block codes over AWGN
channels are known (e.g., \cite{Polt94,Divs}). The bounds on the CEP
are often of the form $\Phi_C(\gamma)=\sum_{h=d}^{n m }\tilde{E}(h)
{F}(\gamma,h)$. It follows that the bit error probability (BEP) is
(e.g., \cite{Ben98, sason00})
\eq{\label{bitcod}\Phi_B(\gamma)=\Phi_C(\gamma) \left|_{\tilde{E}(h)
\Rightarrow \tilde{O}_h}\right.= \sum_{h=d}^{n m }\tilde{O}_h
{F}(\gamma,h).} From
 Th. \ref{wghtbit},
 $\tilde{O}_h=\sum_{w=1}^{m k}
\frac{w}{m k} \tilde{O}(w,h) = \frac{h}{m n}\tilde{E}(h)$.
\section{Multiuser Error Probability}
We consider the case when a systematic RS codeword is shared among
more than user or application, where the $i$th partition of size
$n_i$ is assigned to the $i$th user and the last partition
constitutes of the redundancy symbols. It follows that the $j$th
user's SEP and BEP are, respectively,
\eqar{\label{symcodj}\Phi_S^j(\gamma)
=\Phi_C(\gamma)\left|_{{E(h)}\Rightarrow{O_h^j}}\right.,\\
\label{bitcodj} \Phi_B^j(\gamma)=\Phi_C(\gamma) \left|_{\tilde{E}(h)
\Rightarrow \tilde{O}_h^j}\right.,}} where $O_h^j=\sum_{w=1}^{n_j}
\frac{w}{n_j} O^j(w,h)$, $\tilde{O}^j_h=\sum_{w=1}^{ n_j m}
\frac{w}{m n_j}\tilde{O}^j(w,h)$ and $O^j(w,h)$ is given by
(\ref{Oj}).
\thm{\label{muth} For a systematic linear MDS code, the
unconditional SEP (BEP) of all the users is the same regardless of
the size of the partition assigned to each of them.}

\emph{Proof Idea:} For any two users $i$ and $j$,
$O_h^j=O_h^i=\frac{h}{n} E(h)$, regardless of $n_i$ and $n_j$. For
the average binary case, we also have
$\tilde{O}^j_h=\tilde{O}^i_h=\frac{h}{m n} \tilde{E}(h)$. \hfill \QED
\begin{figure}
\centering
\includegraphics[width=3.5in]{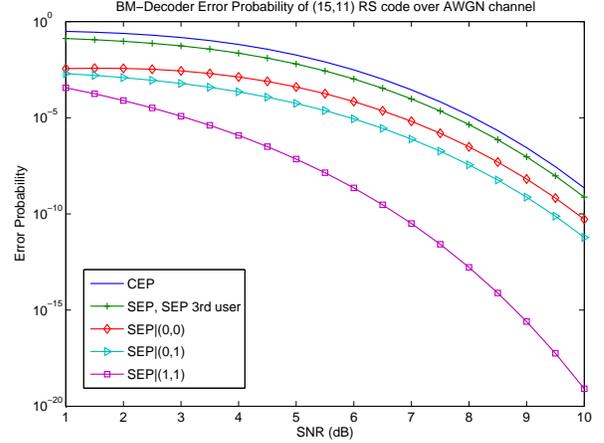}
\caption{\label{BMMu} Multiuser error probability of the BM
decoder.}
\end{figure}

Using the results in this paper,
one could answer interesting questions about the conditional multiuser
error probability. Since the code is linear, we will assume that the all-zero
 codeword is transmitted.
For example, the conditional CEP given that no more than a fraction
$p$ of the $j$th user's symbols are received in error for any
transmitted codeword is given by \footnote{Conditional functions
will have have the same notation as the unconditional ones except
for an underbar.} \eq{\underline{\Phi_C}(\gamma)=\sum_{h=d}^{n}
\sum_{w_j=0}^{\lfloor p n_j \rfloor}
O^j(w_j,h)\sum_{t=0}^{\tau}P_t^h.} (Recall that
$E(h)=\sum_{w_j=0}^{n_j} O^j(w_j,h)$.) Let $O(0,n_j;h) \Def
|\{\mbit{c} \in \mathbb{C}: (\mathcal{W}(\mbit{c})=h) \wedge
(\mathcal{W}(P_i)=0) \wedge (\mathcal{W}(P_j)=n_j)\}|$. The
conditional CEP given that a codeword error results in all $i$th
user's symbols received correctly while all $j$th user's
 symbols received erroneously is given by%
\eq{\underline{\Phi_C}(\gamma)=\sum_{h=d}^{n} O(0,n_j,h)
\sum_{t=0}^{\tau}P_t^h.}

In general for a $p$-partition of the coordinates,
 let $\Omega$ and $\Upsilon$ be the set of users (partitions) whose symbols are all received correctly and
erroneously, respectively, in case of a codeword error. Let $\Delta$
be the set of users with no condition on their error probability.
The conditional error probability is calculated by considering only
the codewords which have a full weight for the coordinates in
$\Upsilon$ and a zero weight for the coordinates in $\Omega$. By
considering only such combinations in the sum of (\ref{par}), the
conditional PWGF
$\underline{\mathbb{P}}(\mathcal{X}_1,\mathcal{X}_2,...,\mathcal{X}_p)$
is derived.

The conditional SEP of the $j$th user is
  \eq{\underline{\Phi_{S}^j}(\gamma) =\Phi_C(\gamma)\left|_{{E(h)} \Rightarrow
  {\underline{O_h^j}}}\right.,}
where $\underline{O_h^j}=\sum_{w=1}^{n_j} \frac{w}{n_j}
\underline{O^j}(w,h)$ and $\underline{O^j}(w,h)$
 is the conditional IOWE of the $j$th partition and is
derived from
$\underline{\mathbb{P}}(\mathcal{X}_1,\mathcal{X}_2,...,\mathcal{X}_p)$
(see (\ref{Oj})).
\begin{figure}
\label{BERfig} \centering
\includegraphics[width=3.5in]{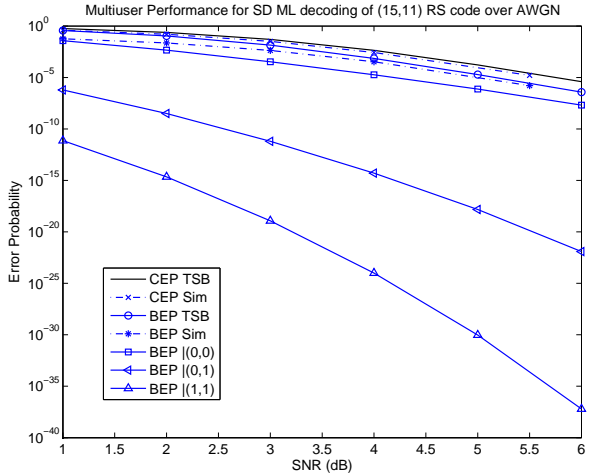}
\caption{\label{SDMu}Multiuser error probability of the SD-ML
decoder}
\end{figure}

 Similarly, for bit-level decoding of the code's binary
image, $\underline{\tilde{O}_h^j}$ will be derived from
$\tilde{\underline{\mathbb{P}}}(\mathcal{X}_1,\mathcal{X}_2,...,\mathcal{X}_p)$.
This conditional binary PWGF only takes into account such codewords
that have a zero weight for the partitions in $\Omega$ and a full binary Hamming
weight for the partitions
in $\Upsilon$.%
The conditional BEP of the $j$th user follows by the substitution
$\tilde{E}(h) \Rightarrow \underline{\tilde{O}_h^j}$ in
(\ref{bitcodj}).
 \ex{\label{HDSEP} Consider an systematic $(15,11)$
RS code and a partition $\mathcal{T}=(3,3,5,4)$ of its coordinates
where the last partition has the redundancy symbols and each of the
first three partitions is assigned to a different user.} Let the RS
code (in fact its binary image) be transmitted over an AWGN channel
and decoded by the Berlekamp-Massey (BM) decoder. In Fig.
\ref{BMMu}, the unconditional CEP and SEP (which by Th. \ref{muth} is equal to
the SEP of the $3$rd user) are plotted. The conditional SEP of the
$3$rd user is plotted for three cases; a codeword error results in user $1$ and $2$ having a SEP
of i) zero (labeled ($0,0$)), ii) zero and one respectively ($0,1$), iii) one
($1,1$).} In Fig. \ref{SDMu}, we consider the case when the decoder
is the soft-decision ML decoder. Using the averaged binary PWE derived in this paper
and the Poltyrev tangential sphere bound \cite{Polt94}, we
calculate the averaged conditional BEP of the third user given the
three cases; BEP of the first and second users are (0,0), (0,1) and
(1,1) respectively in case of a codeword error. The bounds on the unconditional CEP and BEP
are also plotted and are shown to be tight by comparing with the
simulations (for a specific basis representation), `CEP Sim' and
`BEP Sim' respectively.

It is observed, in Fig. \ref{BMMu} and Fig.
\ref{SDMu}, that the conditional SEP or BEP of a specific user decreases as the number of users
receiving erroneous symbols, in case of a codeword error, increases. %
\section{Conclusion}
In this paper, a closed form formula for the partition weight
enumerator of maximum distance separable (MDS) codes is derived.
  The average PWE is derived for the binary image of MDS
 codes defined over a field of characteristic two.
We show that for MDS codes, all the coordinates have the same weight
in the subcode composed of codewords with equal weight. We prove
that a code has this property iff its dual code has this property.
Consequently, it is shown that the first order Reed Muller codes and
the extended Hamming codes have this property. A common
approximation used to evaluate the symbol and bit error
probabilities is shown to be exact for MDS codes. These results are
employed to study the error probability when a Reed Solomon code is
shared among different users and the decoder is either a bounded
minimum distance decoder or a maximum likelihood decoder.

\section*{Acknowledgment} This research was supported by NSF grant
no. CCR-0118670 and grants from Sony, Qualcomm, and the Lee Center
for Advanced Networking.

\bibliographystyle{IEEEtran}
\bibliography{IEEEabrv,Mybibalpha}
\end{document}